\documentclass[times]{speauth}

\usepackage{tikz}
\usetikzlibrary{chains,positioning}
\usepackage{underscore}
\usepackage{todonotes}
\usepackage{algorithm}
\usepackage[noend]{algorithmic}

\usepackage{numcompress}

\makeatletter
\def\ps@pprintTitle{%
  \let\@oddhead\@empty
  \let\@evenhead\@empty
  \let\@oddfoot\@empty
  \let\@evenfoot\@oddfoot
}
\makeatother
\graphicspath{{../figures/}{.}}
\usepackage{booktabs}
\usepackage{paralist}
\usepackage{siunitx}
\usepackage{graphicx}
\usepackage{amsmath}
\usepackage{comment}
\usepackage{url}
\usepackage[T1]{fontenc}
\usepackage{times}
\usepackage{subfig}
\usepackage{mathtools}

\DeclarePairedDelimiter\ceil{\lceil}{\rceil}

\usepackage[colorlinks=false]{hyperref}

\usepackage{url}

\usepackage{breakurl}
\begin{document}

\newcommand{\Censeighten}{\textsc{Census1881}}
\newcommand{\CensInc}{\textsc{CensusInc}}
\newcommand{\Wikileaks}{\textsc{Wikileaks}}
\newcommand{\Weather}{\textsc{Weather}}
\newcommand{\Censtwothousand}{\textsc{Census2000}}

\newcommand{\Censeightensrt}{\textsc{Census1881}$^{\mathrm{sort}}$}
\newcommand{\CensIncsrt}{\textsc{CensusInc}$^{\mathrm{sort}}$}
\newcommand{\Wikileakssrt}{\textsc{Wikileaks}$^{\mathrm{sort}}$}
\newcommand{\Weathersrt}{\textsc{Weather}$^{\mathrm{sort}}$}
\newcommand{\Censtwothousandsrt}{\textsc{Census2000}$^{\mathrm{sort}}$}
\setlength{\tabcolsep}{3pt}

\runningheads{D. Lemire, G.~Ssi-Yan-Kai, O.~Kaser}{Roaring: Consistently faster and smaller compressed bitmaps}

\title{Consistently faster and smaller compressed bitmaps with Roaring}

\author{D. Lemire\affil{1}, G.~Ssi-Yan-Kai\affil{2}, O.~Kaser\affil{3}}

\address{\affilnum{1}LICEF Research Center, TELUQ, Montreal, QC, Canada\break
\affilnum{2}42 Quai Georges Gorse, Boulogne-Billancourt, France\break
\affilnum{3}Computer Science and Applied Statistics,
UNB Saint John, Saint John, NB, Canada
}

\cgsn{Natural Sciences and Engineering Research Council of Canada}{261437}
\corraddr{Daniel Lemire, LICEF Research Center, TELUQ,
Universit\'e du Qu\'ebec,
5800 Saint-Denis,
Office 1105,
Montreal (Quebec),
H2S 3L5 Canada. Email: lemire@gmail.com}

\begin{abstract}
Compressed bitmap indexes are used in
databases and search engines. 
Many bitmap compression techniques have 
been proposed, almost all relying primarily on run-length encoding (RLE).
However,
on unsorted data, we can get superior performance with a hybrid compression technique that uses both uncompressed bitmaps and packed arrays inside a two-level tree.
An instance of this technique, Roaring, has recently been proposed. Due to its good performance, it has been adopted by several production platforms (e.g., Apache Lucene, Apache Spark, Apache Kylin and Druid).

Yet there are cases where run-length encoded bitmaps are smaller than the original Roaring bitmaps---typically when the data is sorted so that the bitmaps contain long compressible runs.
To better handle these cases, we build a new Roaring hybrid that combines uncompressed bitmaps, packed arrays and RLE compressed segments. The result is a new Roaring format that compresses better.

Overall, our new implementation of Roaring can be several times faster (up to two orders of magnitude) than the implementations of traditional RLE-based alternatives (WAH, Concise, EWAH) while compressing better. We review the design choices and optimizations that make these good results possible.
\end{abstract}

\keywords{performance; measurement; index compression; bitmap index}

\maketitle

\section{Introduction}

Sets are a fundamental abstraction in
software. They can be implemented in various
ways, as hash sets, as trees, and so forth.
In databases and search engines, sets are often an integral
part of indexes. For example, we may need to maintain a set
of all documents or rows  (represented by numerical identifiers)
that satisfy some property. Besides adding or removing
elements from the set, we need fast functions
to compute the intersection, the union, the difference between sets, and so on.

To implement a set
of integers, a particularly appealing strategy is the
bitmap (also called bitset or bit vector). Using $n$~bits,
we can represent any set made of the integers from the range
$[0,n)$: it suffices to set the $i^{\mathrm{th}}$~bit to one if integer $i$ is in the set. Commodity processors
use words of $W=32$ or $W=64$~bits. By combining many such words, we can
support large values of $n$. Intersections, unions and differences can then be implemented as bitwise AND, OR and AND~NOT operations.
More complicated set functions can also be implemented as bitwise operations~\cite{SPE:SPE2289}.
When the bitset approach is applicable, it can be orders of
magnitude faster than other possible implementation of a set (e.g., as a hash set) while using several times less memory~\cite{colantonio:2010:ccn:1824821.1824857}.

Unfortunately, conventional bitmaps are only applicable when
the cardinality of the set ($|S|$) is relatively large compared to the
universe size ($n$), e.g., $|S| > n /64$.
They are also suboptimal when the set is made of
consecutive values (e.g., $S= \{1,2,3,4,\ldots,99,100\}$).

One popular approach has been to compress bitmaps
with run-length encoding (RLE). Effectively,
instead of using $\lceil n/W\rceil$~words containing $W$ bits
for all bitmaps, we look for runs of consecutive  words containing only
ones or only zeros, and we replace them
with markers that indicate which value is being repeated,
and how many repetitions there are. This RLE-based approach
was first put into practice by Oracle's BBC~\cite{874730},
using 8-bit words. Starting with WAH~\cite{wu2008breaking},
there have been many variations on this idea, including
Concise~\cite{colantonio:2010:ccn:1824821.1824857},
EWAH~\cite{arxiv:0901.3751}, COMPAX~\cite{netfli}, and so on. (See \S~\ref{sec:relwork}.) For better performance, 
they use wider words than BBC (32-bit or 64-bit words).

When processing such RLE-compressed formats, one may need to read
every compressed word to determine whether a value
is in the set. Moreover, computing the
intersection or union between two bitmaps $B_1$ and $B_2$ has complexity
$O(|B_1|+|B_2|)$ where $|B_1|$ and $|B_2|$ are the compressed
sizes of the bitmaps. This complexity is worse than that of
a hash set, where we can compute an intersection with
an expected-time complexity of $O(\min(|S_1|,|S_2|))$ where $|S_1|$ and $|S_2|$ are the cardinalities of the sets.
Indeed, it suffices to iterate over the smallest sets, and for each value, check whether it is in the larger set.
Similarly, we can compute an in-place union, where the result is stored in the largest hash set,  by inserting all
of the values from the small set in the large set, in expected time $O(\min(|S_1|,|S_2|))$. It is comparatively more difficult
to compute in-place unions with RLE-compressed bitmaps such as Concise or WAH\@. Moreover, checking whether a given value is 
in an RLE-compressed bitmap
may require a complete scan of the entire bitmap in $O(|B|)$ time.
Such a scan can be
hundreds of times slower
than checking for the presence of a value in an uncompressed bitmap or hash map.

 For better performance, Chambi et al.~\cite{SPE:SPE2325} proposed the Roaring bitmap format, and made it available as an open-source library.\footnote{\url{https://github.com/RoaringBitmap/RoaringBitmap}}
Roaring partitions the space $[0,n)$ into \emph{chunks} of  $2^{16}$~integers ($[0,2^{16}),[2^{16},2\times 2^{16}), \ldots$).
Each  set value is stored in a container corresponding to its chunk.
 Roaring  stores  dense and sparse chunks  differently.
Dense chunks (containing more than \num{4096}~integers) are stored using conventional bitmap containers (made of $2^{16}$~bits or \SI{8}{kB})
 whereas sparse chunks use smaller containers made of packed sorted arrays of 16-bit integers.
All integers in a chunk share the same 16~most-significant bits. 
The containers are stored in an array along with the most-significant bits.
Though we refer to a Roaring bitmap as a bitmap, it is a hybrid data structure, combining uncompressed bitmaps with sorted arrays.

Roaring allows fast random access. To check for the presence of a 32-bit integer $x$, we seek
the container corresponding to the 16~most significant bits of $x$, using a binary search. If a bitmap container is found, we check the corresponding bit (at index $x \bmod{2^{16}}$); if an array container is found, we use a binary search.
Similarly, we can compute the intersection between two Roaring bitmaps without having to access all of the data. Indeed, suppose that we have a Roaring bitmap $B_1$ containing
only a few values, which all fall
in the range $[0,2^{16})$.  This implies it uses an array container.
Suppose we have another Roaring bitmap $B_2$ over the same range but containing many values; it
can only use
a bitmap container. In that case, computing the intersection between $B_1$ and $B_2$ can be done in time $O(|B_1|)$, since it suffices to iterate over the set values of $B_1$ and check whether the corresponding bit is set in the bitmap container of $B_2$. Moreover, to compute the in-place union of $B_1$ and $B_2$, where the result is stored in the large bitmap ($B_2$), it suffices to iterate through the values of $B_1$ and set the corresponding bits in $B_2$ to 1, in time $O(|B_1|)$.

Fast intersection, union and difference operations are made possible through fast operations between containers, even when these containers have different types (i.e., bitmap-vs-bitmap, array-vs-array, array-vs-bitmap, bitmap-vs-array). Though conceptually simple, these operations between containers must produce new containers that are either arrays or bitmap containers. Because converting between container types may be expensive, we found it useful to predict
the container type as part of the computation. For example, if we must compute the union between two array containers such that the sum of their cardinalities exceeds \num{4096}, we preemptively create a bitmap container and store the result of the union. Only if the resulting cardinality falls below \num{4096} do we convert the result back to an array container. (See \S~\ref{sec:logicalOp} for more details.)

We found that Roaring could compress better than WAH and Concise while providing superior speed (up to two orders of magnitude). Consequently, many widely used systems have since adopted Roaring: Apache Lucene~\cite{RoaringDocIdSet,RoaringDocIdSetBlogPost} and its extensions (Solr, Elastic), Apache Spark~\cite{Zaharia:2010:SCC:1863103.1863113},
 Apache Kylin, Druid~\cite{Yang:2014:DRA:2588555.2595631}, and so forth.

However, the original Roaring has a limitation in some scenarios because it does not compress long runs of values.  Indeed, given a bitset made of a few long runs (e.g., all integers in $[10,1000]$), Roaring---as presented so far---can only offer suboptimal compression. This was reported in Chambi et al.~\cite{SPE:SPE2325}: in one dataset, Concise and WAH offer better compression than Roaring (by about \SI{30}{\percent}). Practitioners working with
Druid also reported  that  Roaring could use more space (e.g., \SI{20}{\percent} or more) than Concise. As previously reported~\cite{SPE:SPE2325}, even when Roaring offers suboptimal compression, it can still be expected to be faster for many important operations (see \S~\ref{sec:perfheap}). Nevertheless,
there are cases where storing the information as runs  both reduces memory usage drastically and accelerates the computation. If we consider the case of a bitmap made of all integers in $[10,1000]$,
Roaring without support for runs would use \SI{8}{kB}, whereas a few bytes ought to suffice. Such unnecessarily large bitmaps can stress memory bandwidth. Moreover, computing the intersection of two bitmaps representing the ranges
$[10,1000]$ and $[500,10000]$ can be done in a few cycles when using RLE-compressed bitmaps, but the original Roaring
 would require intersecting two bitmap containers and possibly thousands of cycles. Thus, there are cases where bitmaps without support for run compression are clearly at a disadvantage.

To solve this problem, we decided to add a third type of container to Roaring, one that is ideally suited to coding data made of runs of consecutive values. The new container is conceptually simple: given a run (e.g.,  $[10,1000]$), we  store the starting point ($10$) and its length minus one (990).
By packing the starting points and the lengths in pairs, using 16~bits each, we preserve the ability to support fast random access by
binary search
through the coded runs. We refer to the new format as Roaring+Run when it needs to be  distinguished from the original Roaring.

Adding a third container type introduces several engineering problems, however. For one thing, instead of a handful of possible container-type interactions~(bitmap-bitmap, array-array and array-bitmap), we have about twice as many~(bitmap-bitmap, array-array, array-bitmap, run-bitmap, run-array and run-run).
Since
we need to predict the container type as part of the computation to avoid expensive container conversions, new heuristics are needed.
It is also not a priori clear whether introducing a new container type could comprehensively solve our compression issues.
Moreover, assuming that a new Roaring format improves compression, is it at the expense of speed?

Thankfully, we are able to successfully implement
a new Roaring model, made of three container types, that is superior in almost every way to WAH, Concise and EWAH in all our tests---including cases where the original Roaring performs more poorly. Compared to the original Roaring, Roaring+Run can improve the compression ratios by up to an order of magnitude. In what might be our worst-case scenario (\Censeighten{}), the new version has half the speed while saving only about 5\% of storage: even in this case, the new version remains faster by one or two orders of magnitude
when
compared to an implementation of WAH and Concise. Roaring+Run is consistently faster and smaller than popular RLE-based compression schemes (WAH,  Concise and EWAH) in our tests.

\section{Related work}\label{sec:relwork}

There are many RLE-based compression formats. For example, WAH organizes the data in
literal and fill words.
Literal words contain a mix of $W-1$ zeros and ones (e.g., $01011\cdots01$)
where $W$ denotes the word size in bits: typically $W=32$ or $W=64$.
Fill words are made of just $W-1$~ones or just $W-1$~zeros (i.e.,
$11\cdots11$
or  $00\cdots00$).
WAH compresses sequences of consecutive identical  fill words.
The most significant bit of each word distinguishes between
fill and literal words. When it is set to one, the remaining $W-1$~bits store the $W-1$~bits of a literal word. When it is set to zero, the second most significant bit indicates the bit value whereas the remaining bits are used to store the number of consecutive identical fill words (the run length).
Concise is a variation that reduces the memory usage when the bitmap is
moderately
 sparse~\cite{colantonio:2010:ccn:1824821.1824857}. Instead of storing the run length
 using $W-2$~bits, Concise uses only $W-2- \ceil*{\log_2 (W)}$~bits to indicate a run length $r$, reserving $\ceil*{\log_2 (W)}$~bits to store a  value $p$.
 When $p$ is non-zero, we decode $r$~fill words, plus a single $W-1$~bit word with its $p^{\mathrm{th}}$~bit flipped.
 Whereas WAH would use 64~bits per value to store the set $\{0,62, 124,\ldots\}$, Concise would only use 32~bits per value.

 EWAH is similar to WAH except that
 it uses a marker word that indicates the number of fill words to follow, their type, as well as the number of literal words to follow.
 Unlike WAH and Concise, which represent the bitmap as a series of $W-1$-bit words, EWAH uses $W$-bit words.
 The EWAH format~\cite{arxiv:0901.3751} supports a limited form of skipping because
it uses marker words  to record the length of the sequences of
fill and literal words.
For EWAH,
if there are long sequences of
literal words, one does not need to access them all
when
seeking data that is further along.  Guzun et
al.~\cite{guzuntunable,Guzun2015} found that EWAH offers better speed than WAH and Concise, and our own experiments support this observation (see \S~\ref{sec:exp}). 

Beside WAH, Concise and EWAH, there are many other similar alternatives.  We refer the interested reader to Chen et al.~\cite{7040519}, who review over a dozen compressed bitmap formats.

The general idea behind Roaring---using different container types depending on the data characteristics---is not novel. It is similar to O'Neil and O'Neil's RIDBit external-memory system: a B-tree of  bitmaps, where  a list is used instead when the density of a chunk is too small~\cite{4318091,Rinfret:2001:BIA:375663.375669}.  Similarly,
Culpepper and Moffat proposed a hybrid inverted index (\textsc{hyb+m2}) where some sets of document identifiers are stored as compressed arrays whereas others are stored as bitmaps~\cite{culpepper:2010:esi:1877766.1877767}.
To reduce cache misses, Lemire et al.\ partitioned  the Culpepper-Moffat index into chunks that fit in L3~cache~\cite{simdcompandinter}, effectively creating sets of bitmaps and arrays.
Roaring is also reminiscent of C-store where different columns are stored in different formats depending on their data characteristics (number of runs, number of distinct values)~\cite{1083658}.
It is likely that we could find many other similar instances.

\section{Application context}\label{sec:application}

Sets can be used for many purposes. We are interested
in applications that use sets of integers as part of an index.
For example,
one might index an attribute in a database or a word in a set
of documents: for each attribute value,
we have a set of numerical record identifiers.

Indexes are most useful when there are many record identifiers.
We expect the integer values in the set to span
a wide range of values, i.e., 
at least hundreds of thousands.
We are interested in cases where bitmaps are likely applicable: on average, there should be more than a few dozen integer values per set.

 Though we can expect updates, we assume that
most of the processing is spent answering queries that do not require
modifying the set.
Thus, our application setting is one that might be described as
analytical as opposed to transactional.
If we can assume that sets are immutable in the normal course
of the application, this has the added benefit of
simplifying parallelisation and concurrency. Moreover,
the sets can be stored on disk and memory mapped on demand.

We need to provide functions that enable the creation of the
sets,
as well as their
serialization (for later reuse). Otherwise,
 we are most interested by the
union and intersection between two or more sets, as a new set.
It is not uncommon to need to process many more than two sets at once.
We are also interested in random-value accesses, e.g., checking whether a
 value is contained in a set.

\section{Roaring bitmap}\label{sec:roaring}

For a detailed presentation of the original Roaring model, we refer the interested reader to Chambi et al.~\cite{SPE:SPE2325}. We summarize the main points and focus on the new algorithms and their implementations.

Roaring bitmaps are used to represent sets of 32-bit unsigned integers. At a high level, a Roaring bitmap implementation is a key-value data structure  where each key-value pair represents the set $S$ of all 32-bit integers that share the same most significant 16~bits. The key is made of the shared 16~bits, whereas
the value is a container storing
the remaining 16~least significant bits for each member of $S$.
No container ever uses much more than \SI{8}{kB} of memory. Thus, several such small containers fit in the L1 CPU cache of most processors: the last Intel desktop processor to have less than   \SI{64}{kB} of total (data and code) L1 cache was the P6 created in 1995,
whereas most mobile processors have \SI{32}{kB} (e.g., NVidia, Qualcomm) or \SI{64}{kB} (e.g., Apple) of total L1 cache.

 In our actual implementation, the key-value store is implemented as two arrays: an array of packed 16-bit values representing the keys and   an array of containers. 
The arrays expand dynamically in a standard manner when there are insertions. Alternatively, we could use a tree structure for faster insertions, but we expect Roaring bitmaps to be immutable for most of the life of an application. An array minimizes storage. 

In a system such as Druid, the bitmaps are created, stored on disk and then memory-mapped as needed. When we serialize the bitmaps,  we interleave with the 16-bit keys, the cardinalities of the corresponding containers: cardinalities are stored as 16-bit values (indicating the cardinality minus one). 
If needed, we also use an uncompressed bitmap containing at least one bit per container to indicate whether the corresponding container is a run container. 

The structure of each container is straightforward (by design):
\begin{itemize}
\item A bitmap container is an object made of \num{1024}~64-bit words (using \SI{8}{kB}) representing an uncompressed bitmap, able to store all sets of 16-bit integers. The container can be serialized as an array of 64-bit words. We also
maintain
a counter to record how many bits are set to 1.

In some cases, the range of values might not cover the full range $[0,2^{16})$ and a smaller bitmap might be sufficient---thus improving compression. However, the bitmap containers would then need to grow and shrink dynamically. For simplicity, we use fixed-size bitmap containers.

Counting the number of 1-bits in a word can be relatively expensive if done na\"ively, but modern processors have bit-count instructions---such as \texttt{popcnt} for x64 processors and \texttt{cnt} for the 64-bit ARM architecture---that can do this count using sometimes as little as a single clock cycle. According to our tests, using dedicated processor instructions can be several times faster than using either tabulation or other conventional alternatives~\cite{warr:hackers-delight-2e}.
Henceforth, we refer to such a function as \texttt{bitCount}: it is provided in Java as the \texttt{Long.bitCount} intrinsic. We assume that the platform has a fast  \texttt{bitCount} function.
\item An array container is an object containing a counter keeping track of the number of integers, followed by a packed array of sorted 16-bit unsigned integers. It can be serialized as
an array of
16-bit values.

We implement array containers as dynamic arrays that grow their capacity
using a standard approach. That is, we keep a count of the used entries in an underlying array that has typically some excess capacity. When the array needs to grow beyond its capacity, we allocate a larger array and copy the data to this new array. Our allocation heuristic is as follow: when the capacity is small (fewer than 64~entries), we double the capacity;
when the capacity is moderate (between 64 and 1067~entries), we multiply the capacity by $3/2$;
when the capacity is large (1067~entries and more), we multiply the capacity by $5/4$.
Furthermore, we never allocate more than the maximum needed (\num{4096}) and if we are within one sixteenth of the
 maximum ($>3840$), then we allocate the maximum right away (\num{4096})
 to avoid any future reallocation. A simpler heuristic where we double the capacity whenever it is insufficient would be faster, but it would allocate on average (over all possible sizes) a capacity that exceeds the size by \SI{50}{\percent}
 whereas the capacity exceeds the size by only \SI{13}{\percent} in our model. In this sense, we trade speed for reduced memory usage.
When the array container is no longer expected to grow, the programmer can use a \texttt{trim} function to copy the data to a new array with no excess capacity.
\item Our new addition, the run container, is made of a packed array of pairs of 16-bit integers. The first value of each pair represents a starting value, whereas the second value is the length of a run.
For example, we would store the values $ 11,12,13,14,15$ as the pair $11,4$ where 4 means that beyond 11 itself, there are 4 contiguous values that follow. In addition to this packed array, we need to maintain the  number of runs stored in the packed array. Like the array container, the run container is stored in a dynamic array. 
During serialization, we write out
the number of runs,
followed by the corresponding packed array.

Unlike
an array or bitmap container, a run container does not keep track of its cardinality; its
cardinality can be computed on the fly by summing
the lengths of the runs. In most applications, we expect the number of runs to be often small: 
the computation of the cardinality should not be a bottleneck.

However, as part of the serialization process, the cardinality
of the run container \emph{is} computed and stored. Hence, if we access the Roaring+Run  bitmaps in their serialized form (as memory-mapped bitmaps), the cardinality of run containers is pre-computed.
\end{itemize}

When starting from an empty Roaring  bitmap, if a value is added, an array container is created.
When inserting a new value in an array container, if the cardinality exceeds \num{4096},
then the container is transformed into a bitmap container. 
On the other hand%
, if a value is removed from a bitmap container so that its size falls to \num{4096}~integers,
 then it is transformed into an array container. Whenever a container becomes empty, it is removed
from the top-level key-value structure
along with the corresponding key.

Thus, when first creating a Roaring  bitmap, it is usually made of array and bitmap containers. Runs are not compressed.
Upon request,
the storage of the Roaring 
bitmap can
be optimized using the \texttt{runOptimize} function.
This triggers a scan
through the array and bitmap containers
that converts them, if helpful,
to run containers. In a given application, this might be done
prior to storing the bitmaps as immutable objects to be queried.
Run containers may also arise
from calling
a function to   add a range of values.

To decide the best container type, we are motivated to minimize storage.
In serialized form, a run container uses $2+4r$~bytes given $r$~runs, a bitmap container always uses \num{8192}~bytes and an array container uses $2c+2$~bytes, where $c$ is the cardinality.
Therefore, we apply the following rules:
\begin{itemize}
\item All array containers are such that they use no more space than they would as a bitmap container: they contain no more  than \num{4096}~values.
\item Bitmap containers use less space than they would as array containers: they contain more than \num{4096}~values.
\item A run container is only allowed to exist if it is smaller than either the array container or the bitmap container that could equivalently store the same values.  If the run container has cardinality greater than \num{4096}~values, then it must contain no more than $\lceil (8192-2)/4 \rceil = 2047$~runs. If the run container has cardinality no more than \num{4096}, then the number of runs must be less than half the cardinality.
\end{itemize}

\paragraph{Counting the number of runs}

A critical step in deciding whether an array or bitmap container should be converted to a run container is to count the number of runs of consecutive numbers it contains. For array containers, we count  this number by iterating through the 16-bit integers and comparing them two by two in a straightforward manner.
   Because array containers
have at most 4096 integers,
this computation is expected to be fast. For bitmap containers,
Algorithm~\ref{algo:countruns} shows how to compute the number of runs.
We can illustrate the core operation of the algorithm using a single 32-bit word containing
6~runs of consecutive ones:
\begin{eqnarray*}
C_i &=&\texttt{000111101111001011111011111000001},\\
C_i\ll{} 1&=&\texttt{001111011110010111110111110000010},\\
(C_i\ll{} 1) \mathrm{~AND NOT~} C_{i} &=&\texttt{001000010000010100000100000000010}.
\end{eqnarray*}
We can verify that $\mathrm{bitCount}(
(C_i\ll{} 1) \mathrm{~AND NOT~} C_{i} )=6$, that is, we have effectively computed the
number of runs. In the case where a run continues up to the left-most bit, and does not continue in the next word, it does not get counted, but we add another term ($(C_i\gg{}63) \mathrm{~AND NOT~} C_{i+1}$ when using 64-bit words) to check for this case.
We use only a few instructions for each word.
Nevertheless, the computation may be expensive---exceeding
the cost of computing the union or intersection between two bitmap containers.
Thus, instead of always computing the number of runs exactly, 
we rely on the observation that no bitmap container with more than 2047 runs should be converted.
As soon as we can produce a lower bound exceeding 2047 on the number of runs, we can stop.
An exact computation of the number of runs is important only when our lower bound is less than 2048.
We found that a good heuristic is to compute the number of runs in blocks of 128~words using a function inspired by Algorithm~\ref{algo:countruns}. We proceed block by block. As soon as the number of runs exceeds the threshold, we conclude that converting
to a run container is counterproductive and abort the computation of the number of runs.
We could also have applied the optimization to array containers as well, stopping the count of the number of runs at  \num{2047}, but this optimization is likely less useful because array containers have
small cardinality compared to bitmap containers.
 A further possible optimization is to omit the last term
from the sum in line~5 of Algorithm~\ref{algo:countruns}, thus underestimating the number of runs, typically by a few percent, but
by up to 1023 in the worst case. 
Computing this lower bound is nearly twice as fast as computing
the exact count in our tests using a recent Intel processor (Haswell microarchitecture).

\begin{algorithm}
\caption{\label{algo:countruns}Routine to compute the number of runs in a bitmap. The left and right shift operators ($\ll{}$ and $\gg{}$) move all bits in a word by the specified number of bits, shifting in zeros. By convention $C\gg{}63$ is the value (1 or 0) of the last bit of the word. We use the bitwise AND~NOT operator.
 }
\centering
\begin{algorithmic}[1]
\STATE \textbf{input}: bitmap $B$ as an array \num{1024}~64-bit integers, $C_1$ to $C_{1024}$.
\STATE \textbf{output}: the number of runs $r$

\STATE $r \leftarrow 0$
\FOR {$i \in \{1,2,\ldots, 1023\}$}
\STATE $r \leftarrow r + \mathrm{bitCount}(
(C_i\ll{} 1) \mathrm{~AND NOT~} C_{i} ) $ \\
\hspace{1cm}
$+
(C_i\gg{}63) \mathrm{~AND NOT~} C_{i+1}$
\ENDFOR
\STATE  $r \leftarrow r + \mathrm{bitCount}(
(C_{1024}\ll{} 1) \mathrm{~AND NOT~} C_{1024} ) $ \\
\hspace{1cm}
$
+ C_{1024}\gg{}63
$
\RETURN $r$
\end{algorithmic}
\end{algorithm}

Efficient conversions between containers are generally straightforward, except for conversions from a bitmap container to another container type. Converting from a bitmap to an array container is reviewed in Chambi et al.~\cite{SPE:SPE2325}. As for the conversion to run containers,
we use Algorithm~\ref{algo:extract} to extract runs from a bitmap.
It is efficient as long as locating the least significant 1-bit in a word is fast.
Thankfully, recent processors include fast instructions to find the index of
the least significant 1-bit in a word or, equivalently, of the number
of trailing zeros (\texttt{bsf} or \texttt{tzcnt} on x64 processors,
\texttt{rbit}  followed by \texttt{clz} on ARM processors). They are accessible in Java through  the \texttt{Long.numberOfTrailingZeroes} intrinsic.
To locate the index of the least significant 0-bit, we  negate the word, and then seek the least significant 1-bit. Otherwise the algorithm relies on inexpensive bit-manipulation techniques.

\begin{algorithm}
\caption{\label{algo:extract}Algorithm to convert the set bits in a bitmap into a list of runs. We assume two-complement's 64-bit arithmetic.
 We use the bitwise AND and OR operations.
}
\centering
\begin{algorithmic}[1]
\STATE \textbf{input}:  a bitmap $B$, as an array of 64-bit words $C_1$ to $C_{1024}$
\STATE \textbf{output}: an array $S$ containing runs of  1-bits found in the bitmap $B$
\STATE Let $S$ be an initially empty list
\STATE Let $i \leftarrow 1$
\STATE $T \leftarrow C_1$

\WHILE {$i \leq 1024$}
\IF {$T=0$}
\STATE $i \leftarrow i+1$
\STATE $T \leftarrow C_i$
\ELSE
\STATE $j \leftarrow $ index of least significant 1-bit in $T$ ($j\in [0,64)$)
\STATE $x \leftarrow j + 64 \times (i - 1)$
\STATE $T \leftarrow  T \mathrm{~OR~} (T - 1)$ \hfill \COMMENT{all bits with indexes $<j$ are set to 1}
\WHILE {$i + 1\leq 1024$ and $T = \texttt{0xFFFFFFFFFFFFFFFF}$}
\STATE $i \leftarrow i+1$
\STATE $T \leftarrow C_i$
\ENDWHILE
\IF {$T = \texttt{0xFFFFFFFFFFFFFFFF}$}
\STATE  $y \leftarrow 64 \times i = 65536$ \hfill \COMMENT{we have $i=1024$}
\ELSE
\STATE $k \leftarrow $ index of least significant 0-bit in $T$  ($k\in [0,64)$)
\STATE $y \leftarrow k + 64 \times (i-1)$
\ENDIF
\STATE append to $S$ a run that goes from $x$ (inclusively) to $y$ (exclusively)
\STATE $T \leftarrow T \mathrm{~AND~}  T+1$  \hfill \COMMENT{all bits with indexes $<k$ are set to 0}

\ENDIF
\ENDWHILE
\RETURN $S$
\end{algorithmic}
\end{algorithm}

\section{Logical operations}\label{sec:logicalOp}

\subsection{Union and intersection}

There are many necessary logical operations, but we present primarily the union and intersection.  They are the most often used, and the most likely operations to
cause performance bottlenecks.

An important algorithm for our purposes is the galloping intersection  (also called exponential intersection) to compute the intersection
between two sorted arrays of sizes $c_1,c_2$. It has  complexity $O(\min(c_1,c_2) \log \max(c_1,c_2))$~\cite{bentley1976almost}. In this  approach,  we pick the next available integer $i$ from the smaller array and seek an integer at least as big in the larger array, looking first at the next available value, then looking twice as far, and so on, until we find an integer that is
not smaller than $i$.
We then use a binary search in the larger array to find the exact location of the first integer not lower than $i$.
We call this process a  galloping search, and repeat it with each value from the smaller array.

A galloping search makes repeated random accesses in a container,
and it could therefore cause expensive cache misses. However, in our case, the potential problem is mitigated by the fact that
all our containers fit in CPU cache.

Intersections between two input Roaring bitmaps start by visiting
the keys from both bitmaps, starting from the beginning. If a key
is found in both
input bitmaps, the corresponding containers are intersected and the result (if non-empty) is added to the output. Otherwise, we advance in the bitmap corresponding to the smallest key, up to the next key that is no smaller than the key of the other bitmap,
using galloping search. When one bitmap runs out of keys, the intersection terminates.

Unions between Roaring data structures
are handled in the conventional manner:
we iterate through the keys in sorted order; if a key is in both
input Roaring bitmaps, we merge the two containers, add the result to the output and advance in the two bitmaps. Otherwise, we clone the container corresponding to the smaller key, add it to the output and advance in this bitmap. When one bitmap runs out of keys, we  append all the remaining content of the other bitmap to the output.

Though we do not use this technique, instead of cloning the containers during unions, we could use a copy-on-write approach whereas a reference to container is stored and used, and a copy is only made if an attempt is made to modify the container further.
This approach can be implemented by adding a bit vector containing one bit per container. Initially, this bit is set to 0, but when the container cannot be safely modified without making a copy, it is set to 1. Each time the container needs to be modified, this bit needs to be checked.
 Whether the copy-on-write approach is worth the added complexity is a subject for future study. However, container cloning was never found  to  a significant computational bottleneck in the course of our development. It is also not clear whether there are applications where it would lead to substantial reduction of the memory usage.
 In any case, merely copying a container in memory can be several times faster than computing the union between two containers:
copying containers is unlikely to be a major bottleneck.

We first briefly review the logical operations between bitmap and array containers, referring the reader to Chambi et al.~\cite{SPE:SPE2325} for algorithmic details.
\begin{description}
\item [Bitmap vs Bitmap:] To compute the intersection between two bitmaps, we first
compute the cardinality of the result using the \texttt{bitCount} function over the bitwise AND of the corresponding pairs of words. If the intersection exceeds \num{4096}, we materialize a bitmap container  by recomputing the bitwise AND between the words and storing them in a new bitmap container. Otherwise, we generate a new array container by, once again, recomputing the bitwise ANDs, and iterating over their 1-bits.
 We find it important  to first determine the right
container type as, otherwise, we would sometimes generate the wrong container and then have to convert it---an expensive process. The performance of the intersection operation between two bitmaps depends crucially on the performance of the \texttt{bitCount} function.

A union between two bitmap containers is straightforward: we execute the bitwise OR between all pairs of corresponding words. There are \num{1024}~words in each container, so \num{1024}~bitwise OR operations are needed. At the same time, we compute the cardinality of the result using the \texttt{bitCount} function on the generated words.

\item [Bitmap vs Array:] The intersection between an array and a bitmap container  can be computed quickly: we iterate over the values in the array container, checking the presence of each 16-bit integer in the bitmap container and generating a new array container that has as much capacity as the input array container. The running time of this operation depends on the cardinality of the array container.  Unions are also efficient: we create a copy of the bitmap and  iterate over the array, setting the corresponding bits.

\item [Array vs Array:] The intersection between two array containers is always a new array container. We allocate a new array container that has its capacity set to the minimum of the cardinalities of the input arrays.  When the two input array containers have similar cardinalities $c_1$ and $c_2$ ($c_1/64 < c_2 < 64 c_1$),
 we use a straightforward merge algorithm with algorithmic complexity $O(c_1 + c_2)$, otherwise we use a galloping intersection  with complexity $O(\min(c_1,c_2) \log \max(c_1,c_2))$~\cite{bentley1976almost}. We arrived at this threshold ($c_1/64 < c_2 < 64 c_1$) empirically as a reasonable choice, but it has not been finely tuned.

	For unions, if the sum of the cardinalities of the array containers is  \num{4096} or less, we  merge the two sorted arrays into a new array container that has its capacity set to the sum of the cardinalities of the input arrays.
	 Otherwise, we generate an initially empty bitmap container. Though we cannot know whether the result will be a bitmap container (i.e., whether the cardinality is larger than \num{4096}), as a heuristic, we suppose that it will be so.
Iterating through the values of both arrays, we set the corresponding bits in the bitmap to 1. Using the \texttt{bitCount} function, we compute cardinality, and then convert the bitmap into an array container if the cardinality is
at most
\num{4096}.

Alternatively, we could be more conservative and predict the cardinality of the union on the assumption that the two containers have independently distributed values over the whole chunk range ($2^{16}$~values). Besides making the code slightly more complicated,
 it is not likely to change the performance characteristics, as the na\"ive model is close to an independence-based model.  Indeed, under such a model the expected cardinality of the intersection would be $\frac{c_1}{2^{16}} \times \frac{c_2}{2^{16}} \times 2^{16} = \frac{c_1 c_2}{ 2^{16}}$.
 The expected cardinality of the union would be $c_1 + c_2 -\frac{c_1 c_2}{ 2^{16}} $. The maximal threshold for an array container is \num{4096}, so we can set  $c_1 + c_2 -\frac{c_1 c_2}{ 2^{16}} =4096$ and
solve for $c_1$ as a function of $c_2$: $c_1 = \frac{2^{16} (4096-c_2)}{2^{16}-c_2}$. In contrast, our simplistic model predicts a cardinality of $c_1+c_2$ and thus a threshold at $c_1 = 4096 - c_2$.
However, since $c_2\leq 4096$, we have that
$\frac{2^{16}}{2^{16}-c_2}= 1 + \frac{c_2}{2^{16}-c_2} \leq  1 + \frac{4096}{2^{16}-4096}= 1.0667$.
That is,  for any fixed value of one container ($c_2$ here), the threshold on the cardinality of the other container,  beyond which we predict a bitmap container, is at most
\SI{6.67}{\percent}
larger under an independence-based estimate, compared with our na\"ive approach.
\end{description}

Given array and bitmap containers, we need to have them interact
with run containers. For this purpose, we introduced several new algorithms and heuristics.

\begin{description}
\item [Run vs Run:] When computing the intersection between
two run containers, we first produce a new run container by a simple intersection algorithm. This new run container has its capacity
set to the sum of the number of runs in both input containers.
 The algorithm starts by considering
the first run, in each container. If they do not overlap, we advance in
the container where the run occurs earlier until they do overlap, or we run out of runs in one of the containers. When we run out of runs in either container, the algorithm terminates. When two runs overlap, we always output their intersection. If the  two runs end at the same value, then we advance in the two run containers. Otherwise, we advance only in the run container that ends first. Once we have computed the answer, after exhausting the runs in at least one container, we check whether the run container should be converted to either a bitmap (if it has too many runs) or to an array container (if its cardinality is too small compared to the number of runs).

The union algorithm is also conceptually simple. We create a new, initially empty, run container that has its capacity set to the sum of the number of runs in both input containers. We iterate over the runs, starting from the first run  in each container. Each time, we pick a run that
has a minimal starting point. We append it to the output
either as a new run, or as an extension of the previous run.
We then advance in the container where we picked the run. Once a container has no more runs, all runs remaining in the other container are appended to the answer. After we have computed the resulting run container,  we convert the run container into a bitmap container if too many runs were created. Checking whether such a conversion is needed is fast, since it can be decided only by checking the number of runs.
There is no need to consider conversion to an array container, because every run present in the original inputs is either present in its
entirety, or as part of an even larger run.  Thus the average run length (essentially our criterion for conversion) is at least as
large as in the input run containers.

\item [Run vs Array:] The intersection between a run container and an array container always outputs an array container. This choice is easily justified: the result of the intersection
has cardinality no larger than the array container, and it cannot contain more runs than
the array container.
We can allocate a new array container that has its capacity set to the
cardinality of the input array container.
Our algorithm is straightforward. We iterate over the values of the array, simultaneously advancing in the run container. Initially, we point at the first value in the array container and the first run in the run container. While the run ends before the array value, we advance in the run container. If the run overlaps the array value, the array value is included in the intersection, otherwise it is omitted.

Determining the best container for storing the  union between a run container and an array is
less straightforward. We could process the run container as if it were an array container, iterating through its integers and re-use our heuristic for the union between two array containers. Unfortunately, this would always result in either an array or bitmap container.
We found that it is often  better
to predict that the
outcome of the union is a run container, and to convert the result to a bitmap container, if we must. Thus, we follow the heuristic for the union between two run containers, effectively treating the array container as a run container where all runs have length one. However, once we have computed the union, we must not only check whether to convert the result to a bitmap container, but also, possibly, to an array container. This check is slightly more expensive, as we must compute the cardinality of the result.

\item [Run vs Bitmap:] The intersection between a run container and a
bitmap container begins by checking the cardinality of the run container. If it is no larger than \num{4096}, then we create an
initially empty array container. We then iterate over all integers
contained in the run container, and check, one by one, whether they
are contained in the bitmap container: when an integer is found to be in the intersection, it is appended to the output in the array container. The running time of this operation is determined by the cardinality of the run container. Otherwise, if the input run container is larger than \num{4096}, then we create a copy of
the input bitmap container. Using fast bitwise operations, we
set to zero all bits corresponding to the complement of the run container  (see Algorithm~\ref{algo:set}). We then check the cardinality of the result, converting to an array container if needed.

The union between a run container and a bitmap container is computed by first cloning the bitmap container. We then set to one all bits corresponding to the integers in the run container, using fast bitwise OR operations  (see again Algorithm~\ref{algo:set}).

\begin{algorithm}[tb]
\caption{\label{algo:set}Algorithm to set a range of bits to 0 or 1 in a bitmap.
}
\centering
\begin{algorithmic}[1]
\STATE \textbf{input}:  a bitmap $B$, as an array of 64-bit words $C_0$ to $C_{1023}$. Integers $i$ and $j$ specifying range $[i,j)$.
\STATE \textbf{output}: the same bitmap with all bits with indexes in $[i,j)$ set to
\begin{itemize}\item 1 if OP is the bitwise OR operation,
\item 0  if OP is the bitwise AND~NOT operation.\end{itemize}
\STATE $ x \leftarrow \lfloor i / 64 \rfloor$
\STATE $y \leftarrow  \lfloor (j-1) / 64 \rfloor$
\STATE $Z \leftarrow \mathrm{0xFFFFFFFFFFFFFFFF}$    \hfill \COMMENT{$Z$ has all its bits set to 1}
\STATE $X \leftarrow Z \ll (i\bmod 64)$ \hfill \COMMENT{$X=\overbrace{11\cdots1}^{64-(i\bmod 64)} \overbrace{00\cdots 0}^{i\bmod 64}$}
\STATE $Y \leftarrow Z \gg (64-(j\bmod 64) \bmod 64)$ \hfill \COMMENT{$Y=\overbrace{00\cdots0}^{64-(j\bmod 64)} \overbrace{11\cdots 1}^{j\bmod 64}$}

\IF {$x=y$}
\STATE $C_x \leftarrow C_x \mathrm{~OP~} (X \mathrm{~AND~} Y)$
\ELSE
\STATE $C_x \leftarrow C_x \mathrm{~OP~} X $
\FOR {$k= x+1, x+2, \ldots, y-1$}
\STATE $C_k \leftarrow C_k \mathrm{~OP~} Z $
\ENDFOR
\STATE $C_y \leftarrow C_y \mathrm{~OP~} Y $
\ENDIF
\STATE return $B$
\end{algorithmic}
\end{algorithm}

In some instances, the result of an intersection
or union between two containers could be most economically
represented as a run container, even if we generate an array or
bitmap container. It is the case when considering
the intersection or union between a run container and a bitmap container.
We could possibly save memory and accelerate later computations by
checking whether the result should be converted to a run container.
However, this would involve keeping track of the number of runs---a
relatively expensive process.

\end{description}

Furthermore, we added the following optimization. Whenever we compute the union between a run container and any other container, we first check whether the run container contains a single run filling up the whole space of 16-bit values ($[0,2^{16})$). In that case, the union must be the other container and we can produce optimized code accordingly. The check itself can be computed in a few inexpensive operations.
This simple optimization accelerates the possibly common case where there are extremely long runs of ones that span multiple containers.

The computation of the intersection is particularly efficient in Roaring when it involves a bitmap container and either an array container or a run container of small cardinality. It is also efficient when intersecting two array containers, with one having small cardinality compared to the other, as we use galloping intersections. Moreover, by design, Roaring can skip over entire chunks of integers, by skipping over keys that are present in only one of the two input bitmaps. Therefore, Roaring is well suited to the problem of intersecting a bitmap having a small cardinality with bitmaps having larger cardinalities. In contrast, RLE-based compression (as in WAH or Concise) offers fewer opportunities to skip input data.

The computation of the union between two Roaring bitmaps is particularly efficient when it involves run containers or array containers being intersected with bitmap containers. Indeed, these computations involve almost no branching and minimal data dependency, and they are therefore likely to be executed efficiently on superscalar processors. RLE-based compression often causes
many data dependencies and much branching.

Another feature of Roaring is that some  of these logical operations can be executed \emph{in place}.  In-place computations avoid unnecessary memory allocations and improve data locality.
\begin{itemize}
\item
The union of a bitmap container with any other container can be written out in the input bitmap container. The intersection between two bitmap containers, or between a bitmap container and some run containers, can also be written out to an input bitmap container.
\item  Though array containers do not support in-place operations, we find it efficient to support in-place unions in a run container with respect to  either another run container or an array container.
In these cases,  it is common that the result of the union is either
smaller or not much larger than combined sizes of the inputs.
The runs in a run container are stored
in a dynamic array that grows as needed
and typically has 
some excess capacity. So we first check whether it would be
possible to store both the input run container and the
output (whose size is bounded by the sum of the inputs).
Otherwise, we allocate the necessary capacity. We then shift
the data corresponding to the input run container from the beginning
of the array to the end. That is, 
if the input run container had $r$~runs, they would be stored in 
the first $4r$~bytes of an array, and we would copy them to the last $4r$~bytes of the array---freeing
the beginning of the array.  We write the result of the union
at the beginning of the array, as usual.
Thus, given enough capacity, this approach enables repeated
unions to the same run container without new memory allocation. We trade the new allocation for a copy within the same array. Since our containers can fit in CPU cache, such a copy can be expected to be fast.
We could, similarly, enable in-place intersections within run or array containers.
 \end{itemize}

A common operation in applications is the aggregation of a long list of bitmaps. When the problem is to compute the intersection of many bitmaps, we can expect a na\"ive algorithm to work well with Roaring: we can  compute the intersection of the first two bitmaps, then intersect the result with the third bitmap, and so forth.
With each new intersection, the result might become smaller, and Roaring
can often efficiently compute the intersection between bitmap having small cardinality and bitmaps having larger cardinalities, as already stated.
Computing the union of many bitmaps requires more care. As already remarked in Chambi et al.~\cite{SPE:SPE2325}, it is wasteful to update the cardinality each and every time when computing the union between several bitmap containers. Though the \texttt{bitCount} function is fast, it can still use a significant fraction of the running time: Chambi et al.~\cite{SPE:SPE2325} report that it reduces the speed by about \SI{30}{\percent}. Instead, we proceed with what we call a ``lazy union''. We compute the union as usual, except that some
unions between containers are handled differently:
\begin{itemize}
\item The union between a bitmap container and any other container type is done as usual, that is, we compute the resulting bitmap container, except that we do not attempt to compute the cardinality of the result. Internally, the cardinality is set to the flag value ``-1'', indicating that the cardinality is currently unknown.
\item When computing the union of a run container with an array container, we always output a run container or, if the number of runs is too great, a bitmap container---even when an array container might be smaller.
\end{itemize}
After the final answer is generated, we ``repair it'' by computing the cardinality of any bitmap container, and by checking whether any run container should be converted to an array container. For even greater speed, we could even make this repair phase optional and skip the computation of the cardinality of the bitmap containers.

We consider two strategies to compute the union of many bitmaps. One approach is a na\"ive two-by-two union: we first compute the union of the first two bitmaps, then the union of the result and the third bitmap and so forth, doing the computation in-place if possible. The benefit of this approach is that we always keep just one intermediate result in memory. In some instances, however, we can
get better results with other algorithms. For example, we can use a heap: put all original bitmaps in a min-heap, and repeatedly poll the two smallest bitmaps, compute their union, and put them back in the heap, as long as the heap contains more than one bitmap. This approach may create many more intermediate bitmaps, but it can also be faster in some instances. To see why that must be the case, consider that the complexity of the union between two bitmaps of size $B$ is $O(B)$, generating a result that might be of size $2B$. Thus, given $N$~bitmaps of size $B$, the na\"ive  approach has complexity $O(B N^2)$, whereas the heap-based approach has complexity $O(B N \log N)$. However, this computational model, favourable to the heap-based approach, does not always apply. Indeed, suppose that the bitmaps are uncompressed bitmaps over the same range of values---as is sometimes the case when Roaring bitmaps are made of bitmap containers. In that case, the computation of a  union between two bitmaps of size $B$ has complexity $O(B)$, and the output has size $B$. We  have that both algorithms,
 na\"ive and heap-based, have the same optimal $O(B N)$ complexity. However, the na\"ive algorithm has storage requirements
in $O(B)$ whereas the heap-based algorithm's storage requirements are in $O(B N)$, indicating that the latter might have worse performance.
We expect that whether one algorithm or the other has better running time is data and format dependent, but in actual application, it might be advantageous to use the na\"ive algorithm if one wishes to have reduced memory usage.

\subsection{Other Operations}

We make available a single software library encompassing both the original Roaring and Roaring+Run (see \S~\ref{sec:software}). We provide optimized implementations of many useful functions.

A logically complete set of operations enables Roaring to be used, via the bit-slicing approach~\cite{253268}, to realize arbitrary Boolean operations.
Thus, our Roaring software supports negation, although it uses the more general \texttt{flip} approach of Java's \texttt{BitSet} class, wherein negation
occurs only within a range.   Besides adding more flexibility, this approach means that there is no need to know the actual universe
size, in the case when the bitset is intended to be over a smaller universe than 0 to $2^{32}-1$.
The \texttt{flip} function in Roaring first determines
the affected containers. The containers found are flipped; those becoming empty are removed.
Missing containers 
that fall entirely within the range
are replaced by
``full'' run containers (a single run from 0 to $2^{16}$). 
When applied to an array container, the flip function uses a binary
search to first determine all values contained in the range. 
We can then determine whether the result of the flip should be an array container or a bitmap container. If the output is an array
container, then the flip can be done in-place, assuming that
there is enough capacity in the container, otherwise a new buffer is allocated. If the output must be a bitmap container, the array
container is converted to a bitmap container and flipped.
Flipping a bitmap container can be done in-place, if needed, using a procedure similar to Algorithm~\ref{algo:set}. In flipping 
a run container, we always first compute the result as a run container. When the container's capacity permits,
an in-place flip avoids memory allocation.  This should be a common situation, because flipping increases the number of runs by at most one.
Thus, there is a capacity problem only when the number of runs increases \emph{and} the original runs fit exactly within the array.
A simple case-based analysis reveals whether the flipped run container requires
an extra run. Suppose we flip the range $[a,b)$ within a container. 
We can prove that the number of runs is increased by one if and only if the following two conditions hold:
\begin{itemize}
\item both $a-1$ and $a$ are contained in the run container, or both values are missing  from the run container,
\item both $b-1$ and $b$ are contained in the run container, or both values are missing  from the run container.
\end{itemize}
After computing the new run container, we check whether it needs to be converted to a bitmap or array container.
Conversion to a bitmap container is rare, as it occurs only when the number of runs has increased from 2047
to 2048.

Although adding negation to intersection and union gives us logical completeness, efficiency is gained by supporting other Boolean
operations directly.  For instance, our Roaring software provides an XOR operation that is frequently useful for
bit-sliced arithmetic\cite{Rinfret:2001:BIA:375663.375669} and that provides symmetric difference
between sets.  Roaring also provides an AND~NOT operation that implements set difference, which is
important in some applications.
The implementation of the symmetric difference is similar to that of the union, with the added difficulty
that the cardinality might be lower than that of
either
of the two inputs.
The implementation of the  difference is similar to that of the intersection.

Our software also supports fast rank and select functions: rank queries count the number of values present in a range whereas select queries seek the $i^{\mathrm{th}}$~value. These queries are accelerated because array and bitmap containers maintain their cardinality as a value that can be quickly queried. 
Moreover, when accessing serialized bitmaps (e.g., through memory-mapped files), the cardinality of all containers is readily available.

Our software also supports the ability to add or remove all values in an interval,  to check efficiently whether two bitmaps intersect (without computing the intersection) and so forth.
We allow users 
to quickly iterate over the values contained in a Roaring bitmap. 
Internally, these iterators are implemented by on-the-fly creation of iterators over containers. We also found it useful to apply 
the flyweight design patterns and to allow programmers to reuse iterator objects---to minimize memory allocation~\cite{Gamma:1995:DPE:186897}.

\section{Experiments}\label{sec:exp}

To validate our results, we present a range of experiments
on realistic datasets.
We focus on bitmap formats: we refer the interested reader
to other work~\cite{colantonio:2010:ccn:1824821.1824857} for
comparisons between bitmaps and other implementations of
sets (such as hash sets or trees).

\subsection{Hardware}
For benchmarking our algorithms, we use   a Linux server with an Intel i7-4770 processor (\SI{3.4}{GHz}, \SI{32}{kB} of L1 data cache, \SI{256}{kB} of L2 cache per core and \SI{8}{MB} of L3 cache). The server
 has \SI{32}{GB} of RAM (DDR3-1600, double-channel).
 Because there is ample memory, we expect that all disk accesses are buffered.
 We disabled Turbo Boost and configured the processor to always run at its highest clock speed.
All our algorithms and software are single-threaded.

\subsection{Software}
\label{sec:software}
We implemented our software in Java and published it as version~0.5 of the Roaring bitmap open-source library.
This library is used by major database systems like Apache Spark,  Apache Kylin  and Druid~\cite{Yang:2014:DRA:2588555.2595631}.
 We use Oracle's JDK 1.8u60 for  Linux (x64) during benchmarking. Benchmarking Java code can
be difficult since one needs to take into account just-in-time compilation, garbage collection and so on. To help cope with these challenges, we wrote all our benchmarks using Oracle's JMH benchmarking system.\footnote{\url{http://openjdk.java.net/projects/code-tools/jmh/}} After an initial warm-up phase, all tests run five times and we record the average time (using wall-clock timings). In all our tests, the
five times vary
by no more than 2\%, so that the average is representative. 
To ensure reproducibility, we make all our software and test data available.\footnote{\url{https://github.com/RoaringBitmap/RoaringBitmap/tree/master/jmh}} 
The interested reader should be able to reproduce all our results merely by launching a script after downloading our software package.

One of the benefits of using Java is that we have access to several
high-quality implementations of competing bitmap formats. For WAH and Concise, we use Metamarkets' CONCISE library (version 1.3.4).\footnote{\url{https://github.com/metamx/extendedset}} This library has
been used for many years by Metamarkets in their Druid database system~\cite{Yang:2014:DRA:2588555.2595631}. It is derived from the original
software produced by Colantonio and Di Pietro~\cite{colantonio:2010:ccn:1824821.1824857} in their work on Concise. We expect it to
perform adequately as an implementation of the Concise format.
  We also make use of the JavaEWAH library (version~1.0.6).\footnote{\url{https://github.com/lemire/javaewah}} JavaEWAH has been used in production for many years in
systems
such as
Apache Hive.  The library has
also been used
in previous research%
~\cite{SPE:SPE2289,arxiv:0901.3751,Nagendra:2015:EPS:2799368.2699483,Nagarkar:2015:CSH:2824032.2824038}. The JavaEWAH library supports both a 32-bit and a 64-bit version of the EWAH format: we test both.

All these Java libraries enable memory-file mapping, by providing versions where the data storage mechanism is abstracted by a Java \texttt{ByteBuffer}. In Java, from a memory-mapped file, one can
extract a \texttt{ByteBuffer} object, and this can be used to randomly access byte or integer values.
Accessing data through a \texttt{ByteBuffer}  can be slower,  because the virtual machine is less able to optimize data access than when the data is in Java's native arrays.
 However, a \texttt{ByteBuffer} object representing a memory-mapped file uses little of the Java heap. 
Hence, this approach reduces the need for garbage collection and may make it less likely that the virtual machine will run out of memory. Moreover, compared to deserializing a bitmap to Java's memory from disk, memory-file mapping can be much faster.
Memory-file mapping works best
for immutable bitmaps%
: we
create them once, store them to disk and retrieve them as needed.

\subsection{Data}

We used four real datasets from earlier studies of
compressed bitmap indexes~\cite{SPE:SPE2325,LemireKaserGutarra-TODS}.
In one instance, the datasets were taken as-is: we did not sort them prior to indexing. In another instance, we sorted them
lexicographically prior to indexing, with the smallest cardinality column being the primary sort key,
the next-smallest cardinality column being the secondary sort key, and so forth~\cite{arxiv:0901.3751}. The net result is that we have two sets of bitmaps
from each data source: one from the data in its original order and one
from sorted data (e.g., \Censeighten{} and \Censeightensrt{}).
For each dataset, we built a bitmap index and
chose 200~bitmaps, $B_1$ to $B_{200}$,
using stratified sampling
 to control for attribute
cardinalities. We present the basic characteristics
of these bitmaps in Table~\ref{tab:chrac}. All of our datasets are publicly available.
Given a collection of bitmaps, we define the universe size as the smallest value $n$ such that all bitmaps viewed as sets of integers are contained in  $[0,n)$. We also report the average
count (or cardinality) of the bitmaps for each collection.
The ratio of universe size over the average cardinality is indicative of the number of bits per value that an uncompressed bitmap would use.
The datasets \Censeighten{}, \Censeightensrt{}, \Wikileaks{} and \Wikileakssrt{} are especially sparse whereas  \CensInc{} and \CensIncsrt{} are the densest ones.

In Table~\ref{tab:iustat}, we report the fraction of all 199~successive intersections (between $B_i$ and $B_{i+1}$)
that are empty. This fraction is above \SI{90}{\percent} for \Censeighten{}, \Censeightensrt{}, \Wikileaks{} and \Wikileakssrt{}, which is consistent with the fact that the bitmaps are sparser in these datasets. There is no dataset where all successive intersections are empty.
When we compute the union of all 200~bitmaps, we get a bitmap
covering the entire universe for the
\CensInc{}, \CensIncsrt{}, \Weather{} and \Weathersrt{} datasets, which is consistent with the fact that they are relatively dense.

\begin{table}
\caption{\label{tab:chrac}Characteristics of
our realistic datasets}
\centering

\subfloat[Description of the bitmaps
]{%
\begin{tabular}{c|rrrr}
\toprule
& \# bitmaps & universe  & average count & count/size \\
&  & size & per bitmap & ratio \\
        \midrule
\CensInc  & 200 & \num{199523} & \num{34610.1} & 5.8 \\
\CensIncsrt  &  200 & \num{199523} & \num{30464.3} & 6.5 \\
\Censeighten  & 200 & \num{4277806} & \num{5019.3} & 852 \\
\Censeightensrt  & 200 & \num{4277735} & \num{3404.0} & 1257\\
\Weather  &  200 & \num{1015367} & \num{64353.1} & 15.8 \\
\Weathersrt  &  200 & \num{1015367} & \num{80540.5} & 12.6\\
\Wikileaks  & 200 & \num{1353179} & \num{1376.8} & 983 \\
\Wikileakssrt & 200 & \num{1353133} & \num{1440.1} & 940\\
\bottomrule
\end{tabular}
}\\
\subfloat[\label{tab:iustat}Statistics regarding intersections and unions
]{%
\begin{tabular}{c|rrr}
\toprule
& fraction of empty & relative size  \\
& intersections & of the union \\
        \midrule
\CensInc  &  0.23 &  1.0 \\
\CensIncsrt  &  0.24 & 1.0 \\
\Censeighten  & 0.97  & 0.11 \\
\Censeightensrt  & 0.97 & 0.15  \\
\Weather  & 0.36 & 1.0  \\
\Weathersrt  & 0.32 & 1.0  \\
\Wikileaks  & 0.91 &  0.18 \\
\Wikileakssrt & 0.95 & 0.17 \\
\bottomrule
\end{tabular}
}\\
\subfloat[\label{tab:comp}Compressed sizes (bits/int), best results in bold]{\small
\begin{tabular}{c|ccccccc}\toprule
 & Concise  & EWAH 64-bit & EWAH &  Roaring & Roaring+Run & WAH \\ \midrule
\CensInc  & 2.9 & 3.9 & 3.3 &  2.7 & \textbf{2.6}   & 2.9\\
\CensIncsrt  & \textbf{0.55}  & 0.90 & 0.64  & 3.0 & 0.60 & \textbf{0.55}\\
\Censeighten  & 25.6 & 43.8 & 33.8 & 16.0 & \textbf{15.1} & 43.8 \\
\Censeightensrt  & 2.5& 4.6 & 2.9 & 6.1 & \textbf{2.2}  & 2.5  \\
\Weather  & 5.9  & 7.9 & 6.7 & \textbf{5.4} & \textbf{5.4} & 5.9\\
\Weathersrt  & 0.43  & 0.86 & 0.54  & 3.2 & \textbf{0.34} & 0.43\\
\Wikileaks  & 10.2 & 19.5 & 10.9  & 16.5 & \textbf{5.9}& 10.2 \\
\Wikileakssrt & 2.2  & 4.7 & 2.7 & 10.7  & \textbf{1.6} & 2.2\\\bottomrule
\end{tabular}
}

\end{table}

\subsection{Compressed Size}

The compressed sizes are given in Table~\ref{tab:comp} in average bits per integer stored. That is, a bitmap using 100~bytes to store 100~values would use 8~bits per integer. We refer the reader to Appendix~\ref{app:detailedstat} for more detailed statistics regarding container usage and sizes per dataset for Roaring with and without runs.

Among the RLE-based formats (all of them but Roaring and Roaring+Run), the best compression is offered by Concise and WAH, with the worst compression offered by the 64-bit version of EWAH\@. Concise and WAH offer the same compression except on one dataset (\Censeighten) where Concise reduces the space usage by a factor of 1.7.
The 64-bit version of EWAH can use twice the storage of the other schemes (all of which are 32-bit formats). However, if we omit the 64-bit version of EWAH, the space usage of all three formats (Concise, EWAH and WAH) is similar (within 30\%).

The RLE-based formats compress sorted datasets much better than unsorted data, sometimes
an order of magnitude better.
For example, WAH and Concise use 5.9~bits per set value for \Weather{} but only 0.43~bits per set value for
\Weathersrt{}.
 This result is consistent with earlier
 work~\cite{arxiv:0901.3751} showing the importance
 of sorting the data prior to indexing it with RLE-compressed bitmap formats.

Roaring without run compression often offers better compression than the RLE-based schemes. For example, on \Censeighten{}, Roaring uses only 60\% of the space used by Concise.
 Yet, on the sorted datasets, the results are much less positive for Roaring. In one  case (\Weathersrt{}), Concise and WAH use $7.4$~times
less space than
 Roaring (0.43~bits vs. 3.2~bits).

However, once we enable run compression in Roaring (Roaring+Run), the results are again more positive. In fact, there is just one case where  Concise uses less storage (\CensIncsrt{}), and the difference is small (8\%).  In what was previously the worst   case (\Weathersrt{}), Roaring is down to 0.34~bits (from 3.2~bits) vs. 0.43~bits for Concise and WAH\@.

We
summarize the results as follows: when the data has been sorted prior to indexing, Roaring with run compression (Roaring+Run) is as good, and even slightly better, than pure RLE-based schemes. When the data is not sorted prior to indexing, Roaring (with or without run compression) can be far superior
in compression.

\subsection{Serialization and Run Optimization}

An expected typical use of run optimization in Roaring consists of
calling the run optimization function (\texttt{runOptimize}) prior to serializing the
 bitmaps.
Table~\ref{tab:ser} presents the effect of run optimization on serialization speed. In our tests,
 we serialize the bitmaps to a byte array (in memory).
 Though it can sometimes be slightly more expensive to
do run optimization when serializing
(by 5\%), it can sometimes be much faster (by over
 $2\times$) because we have to serialize less data. This suggests that
 the run optimization is efficient, at least compared to the cost of
 serializing the bitmaps.  If we were to serialize the bitmaps to a
slow medium,
the benefit of producing less data
 would only grow.

\begin{table}
\caption{\label{tab:ser}Timings in milliseconds for Roaring serialization (200~bitmaps, best results in bold)}
\centering

\begin{tabular}{c|cc}\toprule
 & serialization  & \texttt{runOptimize}  \\
 & & + serialization \\
  \midrule
\CensInc  & \textbf{8.0} & 8.2\\
\CensIncsrt  &4.0& \textbf{1.7}\\
\Censeighten  & 12 & 12 \\
\Censeightensrt  &  2.6 & \textbf{1.7}\\
\Weather  & \textbf{36} & 38\\
\Weathersrt  & 18 & \textbf{8.2}\\
\Wikileaks  & 2.1 & \textbf{1.2}\\
\Wikileakssrt & 1.7 & \textbf{0.7}\\\bottomrule
\end{tabular}
\end{table}

\subsection{Performance of Queries in the Java Heap}
\label{sec:perfheap}

We begin by reviewing the performance of the various
bitmap formats against a range of queries over in-memory bitmaps.
That is, these bitmaps are stored in the Java heap, as most  other
Java objects are.
The results are presented in Table~\ref{tab:inheap}.
In all tests, the source bitmaps are treated as being immutable:
where applicable, new bitmaps containing the answer to the query
are generated.

Because the absolute performance numbers are difficult to
appreciate on their own, we present relative numbers, normalized against
Roaring with run optimization (Roaring+Run  receives value 1.0). Thus, a value of 2.0 would indicate that some operation takes twice as long as it would take with Roaring+Run.

In Table~\ref{tab:rand}, we access the first, second and third quartile position in the universe, and check the presence of the value, for all 200~bitmaps in the set. We observe a slight
decrease of the performance of Roaring when run optimization is
applied (up to 25\%) in some cases, with an improvement in others
(up to 30\%). These differences do not appear very significant
compared to the difference between the Roaring formats and the RLE-based
formats (WAH, EWAH and Concise). Even if we compare against the fastest among them (64-bit EWAH), Roaring can be two orders of magnitude faster. Against Concise and WAH, Roaring is sometimes nearly three orders of magnitude faster.
These drastic results are easily understood: RLE-compressed bitmaps do not sensibly support random access and require a full scan from the beginning.

We then consider the 199~intersections between successive bitmaps (Table~\ref{tab:si}) and the 199~unions between successive bitmaps (Table~\ref{tab:su}). In these tests, we compute the intersections or unions, and then we check
the cardinality of the result against a pre-computed result---thus helping to  prevent  the compiler from optimizing away some of the computation and verifying that all implementations provide the same result.
 Roaring with run optimization (Roaring+Run) is never slower than the original Roaring (within 5\%), but can be up to twice as fast on sorted datasets. Compared to Concise, Roaring+Run is consistently at least 3.5~times faster, and up to hundreds of times faster at computing intersections. The gap between Concise and Roaring+Run is less impressive for unions (between 1.7~times faster and 43~times faster), but still leaves Roaring+Run with a significant advantage.
In these tests,
apart from two cases, Roaring+Run is at least twice as fast as the RLE-based implementations.

Using Oracle's Java Mission Control tool\footnote{\url{www.oracle.com/technetwork/java/javaseproducts/mission-control/}}, we can identify the functions that used most time during the computation of the intersection. 
For simplicity, we choose the \CensInc{} dataset. For Roaring without runs, \SI{57}{\percent} of the time is spent on intersections between pairs of bitmap containers. Intersections between array and bitmap containers account for an additional 
\SI{30}{\percent}. Most of the rest of the time is spent on intersections between array containers. For Roaring+Run, \SI{57}{\percent} of the time is spent on intersection between array containers, and intersections between bitmap containers account for only \SI{11}{\percent} of the time. Intersections between bitmap and array containers account for \SI{29}{\percent} of the running time, whereas intersections between run and array containers account for about \SI{5}{\percent} of the time.
The picture is simpler regarding unions. Most of the time is used
by unions between array containers: \SI{83}{\percent} for Roaring without runs and \SI{68}{\percent} for Roaring+Run. In both cases, unions between array and bitmap containers use about \SI{12}{\percent} of the running time.

Another important test case is the union of all 200~bitmaps. The results are presented in Tables~\ref{tab:eu}~and~\ref{tab:eupq}
for the na\"ive and priority-queue approaches.
In particular, one column in Table~\ref{tab:eupq} directly
compares the na\"ive approach with the priority queue for Roaring+Run.
We see that there is no clear winner between the  na\"ive approach and the priority queue: for \Wikileaks{}, the priority queue is preferable, but for the non-sorted versions of
\CensInc{} and \Censeighten{}, the  na\"ive approach is better for Roaring. In any case, no matter which approach is used, Roaring+Run
is clearly preferable to the RLE-based formats like Concise, being anywhere from $1.8\times$ to nearly $10\times$ faster.

We observe that the priority queue algorithm is ineffective with the original Roaring, reflecting the fact that as we aggregate many bitmaps, it is unable to benefit from the long runs being created in the intermediate bitmaps. Except for one dataset (\Censeightensrt{}), the na\"ive union algorithm is more effective with the original Roaring since, as the result becomes denser, it creates bitmap containers that can compute following unions in-place. Comparing the original Roaring format (without runs) with the RLE-based formats on sorted datasets, we observe that it is generally at least slightly worse, being up to six times slower (\Weathersrt) than Concise.

When assessing the functions responsible for most of the running time 
when computing unions over 200~bitmaps, we find once more that the
union between array containers is relatively expensive. Indeed, 
unions between array containers account for more than \SI{65}{\percent} of the running time when considering the  \CensInc{} dataset, the na\"ive union algorithm and Roaring without runs.

\begin{table}
\captionsetup[subfloat]{farskip=2pt,captionskip=1pt}
\centering
\caption{\label{tab:inheap}Relative timings (Roaring+Run=1), data in Java heap. For each row, the best result is in bold face. }
\subfloat[\label{tab:rand}Random value access]{\small
\begin{tabular}{c|ccccccc}\toprule
& Concise  & EWAH 64-bit & EWAH  & Roaring & Roaring+Run & WAH \\ \midrule
\CensInc & 160 & 19 & 35 & \textbf{0.75} & 1.0 & 170\\
\CensIncsrt & 33 & 8.0 & 12 & \textbf{0.84} & 1.0 & 35\\
\Censeighten & 870 & 190 & 360 & \textbf{1.0} & \textbf{1.0} & 920\\
\Censeightensrt & 52 & 19 & 24 & 1.1 & \textbf{1.0} & 53\\
\Weather & 600 & 76 & 150 & \textbf{0.78} & 1.0 & 610\\
\Weathersrt & 90 & 35 & 42 & \textbf{0.91} & 1.0 & 98\\
\Wikileaks & 48 & 26 & 29 & 1.1 & \textbf{1.0} & 49\\
\Wikileakssrt & 14 & 9.4 & 9.8 & 1.3 & \textbf{1.0} & 15\\\bottomrule
\end{tabular}
}

\subfloat[\label{tab:si}successive intersections]{\small
\begin{tabular}{c|ccccccc}\toprule
& Concise  & EWAH 64-bit & EWAH  & Roaring & Roaring+Run & WAH \\ \midrule
\CensInc & 5.5 & 2.6 & 3.9 & \textbf{0.99} & 1.0 & 6.4\\
\CensIncsrt & 3.5 & 1.4 & 1.7 & 1.1 & \textbf{1.0} & 3.0\\
\Censeighten & 460 & 94 & 150 & \textbf{0.97} & 1.0 & 370\\
\Censeightensrt & 61 & 19 & 23 & 1.2 & \textbf{1.0} & 55\\
\Weather & 5.5 & 2.1 & 3.3 & \textbf{0.98} & 1.0 & 4.7\\
\Weathersrt & 5.7 & 2.4 & 2.6 & 1.4 & \textbf{1.0} & 5.1\\
\Wikileaks & 7.2 & 3.6 & 3.6 & \textbf{0.96} & 1.0 & 6.9\\
\Wikileakssrt & 9.1 & 5.9 & 5.9 & 1.5 & \textbf{1.0} & 8.5\\\bottomrule
\end{tabular}
}

\subfloat[\label{tab:su}successive unions]{\small
\begin{tabular}{c|ccccccc}\toprule
& Concise  & EWAH 64-bit & EWAH  & Roaring & Roaring+Run & WAH \\ \midrule
\CensInc & 4.6 & 2.4 & 4.0 & \textbf{1.0} & \textbf{1.0} & 4.2\\
\CensIncsrt & 1.7 & 0.99 & 1.4 & \textbf{1.0} & \textbf{1.0} & 1.5\\
\Censeighten & 43 & 22 & 43 & \textbf{0.92} & 1.0 & 38\\
\Censeightensrt & 8.8 & 6.8 & 8.4 & 1.2 & \textbf{1.0} & 7.8\\
\Weather & 4.4 & 2.0 & 3.8 & \textbf{0.93} & 1.0 & 3.9\\
\Weathersrt & 3.4 & 2.5 & 3.1 & 2.1 & \textbf{1.0} & 3.0\\
\Wikileaks & 3.8 & 4.2 & 4.4 & 1.3 & \textbf{1.0} & 3.6\\
\Wikileakssrt & 2.3 & 2.6 & 2.9 & 1.6 & \textbf{1.0} & 2.3\\\bottomrule
\end{tabular}
}

\subfloat[\label{tab:eu}Entire unions (na\"ive)]{\small
\begin{tabular}{c|ccccccc}\toprule
& Concise  & EWAH 64-bit & EWAH  & Roaring & Roaring+Run & WAH \\ \midrule
\CensInc & 9.7 & 2.6 & 1.7 & 2.9 & \textbf{1.0} & 9.2\\
\CensIncsrt & 8.1 & 5.5 & 4.5 & 25 & \textbf{1.0} & 6.9\\
\Censeighten & 22 & 14 & 4.9 & \textbf{0.51} & 1.0 & 20\\
\Censeightensrt & 15 & 16 & 11 & \textbf{0.71} & 1.0 & 13\\
\Weather & 13 & 5.5 & 3.6 & \textbf{0.99} & 1.0 & 11\\
\Weathersrt & 3.4 & 3.1 & 2.4 & 2.0 & \textbf{1.0} & 3.0\\
\Wikileaks & 5.6 & 5.3 & 3.0 & \textbf{0.28} & 1.0 & 5.3\\
\Wikileakssrt & 11 & 11 & 8.7 & 1.2 & \textbf{1.0} & 9.5\\\bottomrule
\end{tabular}
}

\subfloat[\label{tab:eupq}Entire unions (priority queue)]{\small
\begin{tabular}{c|ccccccccc}\toprule
& Concise  & EWAH & EWAH  & Roaring & Roaring & Roaring & Roaring & WAH \\
&   & 64-bit  &   &  na\"ive & pq & +Run na\"i. &+Run pq &  \\\midrule
\CensInc & 3.9 & 1.4 & 1.6 & 0.5 & 1.1 & \textbf{0.16} & 1.0 & 3.5\\
\CensIncsrt & 7.3 & 4.3 & 5.0 & 24 & 51 & \textbf{1.0} & \textbf{1.0} & 6.7\\
\Censeighten & 2.2 & 1.2 & 2.2 & \textbf{0.35} & 0.79 & 0.69 & 1.0 & 1.9\\
\Censeightensrt & 2.5 & 2.0 & 2.3 & 3.6 & 2.5 & 5.4 & \textbf{1.0} & 2.4\\
\Weather & 3.4 & 1.0 & 1.9 & \textbf{0.17} & 0.93 & \textbf{0.17} & 1.0 & 3.1\\
\Weathersrt & 1.8 & 4.2 & 4.5 & 12 & 45 & 5.8 & \textbf{1.0} & 1.8\\
\Wikileaks & 2.6 & 2.2 & 2.6 & \textbf{0.98} & 1.6 & 3.4 & 1.0 & 2.4\\
\Wikileakssrt & 2.7 & 2.5 & 2.6 & 3.1 & 3.4 & 2.5 & \textbf{1.0} & 2.4\\\bottomrule
\end{tabular}
}
\end{table}

\subsection{Memory-Mapped Performance}

Table~\ref{tab:inmemmap} presents the timings of several
tests on memory-mapped bitmaps, using different formats.
We omit the WAH format because none of our libraries support
it in memory-mapped mode: this is of little consequence
given the similarity between the
Concise and WAH formats and results.
In
these tests, the bitmaps are first written
to disk and then mapped in memory using Java's \texttt{ByteBuffer}.
These bitmaps are immutable. In practice, because
we have sufficient memory compared to the size of the datasets,
we expect all queries to be buffered.
When queries require the generation of a new bitmap,
it is created in the Java heap.

The results of the tests are similar to the case where
we use regular bitmaps stored entirely in the Java heap.
Roaring can be dozens or hundreds of times faster in many cases. 

For the random-access test (Table~\ref{tab:inmemmaprand}), we omit
the results for Concise since the library we used does not have
built-in support for random access. We could easily have implemented our own version using the intersection with a bitmap containing a single set bit, but such an approach might not offer the best efficiency. And, in any case, we cannot expect good results from Concise on this test from previous experiments.

For the results pertaining to the union of all 200~bitmaps (Tables~\ref{tab:inmemmapeu} and~\ref{tab:inmemmapeupq}), we included a
special approach marked by a star ($\star$) developed originally for the Druid engine. Unlike the na\"ive or priority-queue approaches, which always combine bitmaps two-by-two, this approach
takes all bitmaps at once, and using a priority queue, merges
the compressed words into a single output. Chambi et al.~\cite{SPE:SPE2325} described a similar approach for Roaring, where the priority queue worked over containers---this approach has been
replaced by the simpler na\"ive algorithm as the default for the Roaring library. For sorted data sources, the
$\star$
approach is always preferable in this instance. But even in these cases, Roaring+Run is always significantly faster. In two cases, Roaring+Run is  over a hundred times faster than Concise. Even if we focus just on sorted data inputs, the priority-queue approach with Roaring+Run can be up to ten times faster than Concise.

We can once again compare the na\"ive and priority-queue approaches for Roaring+Run, this time using the fact that both tables (Tables~\ref{tab:inmemmapeu} and~\ref{tab:inmemmapeupq}) present the same algorithm (Concise$\star$). Unsurprisingly, there is no clear winner. On sorted datasets, the priority queue is better, 
but the opposite is often 
true for the other datasets. We do expect, however, the na\"ive approach to often have lower memory usage, and since it is particularly simple, it
may be %
a good default.

\begin{table}
\captionsetup[subfloat]{farskip=2pt,captionskip=1pt}
\centering
\caption{\label{tab:inmemmap}Memory-mapped relative timings (Roaring+Run=1). For each row, the best result is in bold face.}
\subfloat[\label{tab:inmemmaprand}Random value access]{\small
\begin{tabular}{c|ccccccc}\toprule
& EWAH 64-bit & EWAH  & Roaring & Roaring+Run\\ \midrule
\CensInc & 21 & 33 & \textbf{0.91} & 1.0\\
\CensIncsrt & 9.2 & 11 & \textbf{0.92} & 1.0\\
\Censeighten & 140 & 250 & \textbf{1.0} & \textbf{1.0}\\
\Censeightensrt & 13 & 16 & \textbf{0.97} & 1.0\\
\Weather & 94 & 170 & \textbf{0.96} & 1.0\\
\Weathersrt & 27 & 31 & \textbf{0.88} & 1.0\\
\Wikileaks & 20 & 20 & \textbf{0.93} & 1.0\\
\Wikileakssrt & 6.2 & 6.3 & \textbf{1.0} & \textbf{1.0}\\\bottomrule
\end{tabular}
}

\subfloat[successive intersections]{\small
\begin{tabular}{c|cccccc}\toprule
& Concise  & EWAH 64-bit & EWAH  & Roaring & Roaring+Run  \\ \midrule
\CensInc & 17 & 2.3 & 3.6 & \textbf{1.0} & \textbf{1.0}\\
\CensIncsrt & 6.2 & 1.3 & 1.6 & 1.3 & \textbf{1.0}\\
\Censeighten & 140 & 79 & 140 & \textbf{0.98} & 1.0\\
\Censeightensrt & 19 & 12 & 15 & \textbf{1.0} & \textbf{1.0}\\
\Weather & 13 & 1.8 & 2.9 & \textbf{0.99} & 1.0\\
\Weathersrt & 6.4 & 2.2 & 2.4 & 1.6 & \textbf{1.0}\\
\Wikileaks & 5.8 & 3.3 & 3.4 & \textbf{0.83} & 1.0\\
\Wikileakssrt & 5.3 & 3.4 & 3.4 & 1.2 & \textbf{1.0}\\\bottomrule
\end{tabular}
}

\subfloat[successive unions]{\small
\begin{tabular}{c|cccccc}\toprule
& Concise  & EWAH 64-bit & EWAH  & Roaring & Roaring+Run \\ \midrule
\CensInc & 28 & 2.5 & 4.1 & \textbf{1.0} & \textbf{1.0}\\
\CensIncsrt & 7.8 & 1.1 & 1.5 & \textbf{1.0} & \textbf{1.0}\\
\Censeighten & 210 & 21 & 38 & \textbf{0.9} & 1.0\\
\Censeightensrt & 22 & 3.7 & 4.3 & \textbf{1.0} & \textbf{1.0}\\
\Weather & 27 & 2.2 & 4.2 & \textbf{0.96} & 1.0\\
\Weathersrt & 15 & 2.5 & 3.1 & 1.6 & \textbf{1.0}\\
\Wikileaks & 15 & 3.7 & 3.9 & 1.5 & \textbf{1.0}\\
\Wikileakssrt & 8.5 & 1.8 & 2.1 & 1.3 & \textbf{1.0}\\\bottomrule
\end{tabular}
}

\subfloat[\label{tab:inmemmapeu}Entire unions (na\"ive)]{\small
\begin{tabular}{c|ccccccc}\toprule
& Concise  & Concise$\star$   & EWAH 64-bit & EWAH  & Roaring & Roaring+Run  \\  \midrule
\CensInc & 16 & 110 & 1.7 & 2.8 & 2.3 & \textbf{1.0}\\
\CensIncsrt & 2.7 & 2.1 & \textbf{0.88} & 1.1 & 3.3 & 1.0\\
\Censeighten & 210 & 21 & 4.4 & 13 & \textbf{0.53} & 1.0\\
\Censeightensrt & 100 & 2.2 & 12 & 15 & \textbf{0.73} & 1.0\\
\Weather & 45 & 190 & 3.4 & 5.4 & \textbf{0.96} & 1.0\\
\Weathersrt & 14 & 1.4 & 2.6 & 3.3 & 1.7 & \textbf{1.0}\\
\Wikileaks & 34 & 1.7 & 3.0 & 5.4 & \textbf{0.28} & 1.0\\
\Wikileakssrt & 71 & 2.3 & 9.3 & 11 & 1.1 & \textbf{1.0}\\\bottomrule
\end{tabular}
}

\subfloat[\label{tab:inmemmapeupq}Entire unions (priority queue)]{\small
\begin{tabular}{c|ccccccccc}\toprule
& Concise & Concise & EWAH & EWAH  & Roaring & Roaring & Roaring & Roaring  \\
& PQ & $\star$  & 64-bit  &   &  na\"ive & pq & +Run na\"i. &+Run pq  \\\midrule
\CensInc & 8.0 & 22 & 1.4 & 1.8 & 0.47 & 1.1 & \textbf{0.2} & 1.0\\
\CensIncsrt & 8.2 & 8.0 & 3.1 & 4.0 & 12 & 26 & 3.5 & \textbf{1.0}\\
\Censeighten & 12 & 14 & 1.2 & 2.1 & \textbf{0.34} & 0.68 & 0.65 & 1.0\\
\Censeightensrt & 9.8 & 10 & 1.9 & 1.9 & 3.3 & 1.9 & 4.8 & \textbf{1.0}\\
\Weather & 13 & 34 & 1.0 & 2.0 & \textbf{0.18} & 0.93 & \textbf{0.18} & 1.0\\
\Weathersrt & 4.1 & 4.0 & 2.9 & 3.3 & 5.2 & 12 & 3.0 & \textbf{1.0}\\
\Wikileaks & 12 & 5.7 & 2.3 & 2.6 & \textbf{0.96} & 1.5 & 3.4 & 1.0\\
\Wikileakssrt & 11 & 4.7 & 2.3 & 2.5 & 2.5 & 2.2 & 2.2 & \textbf{1.0}\\\bottomrule
\end{tabular}
}
\end{table}

\section{Conclusion}
\label{sec:conclusion}

We have shown how a hybrid bitmap format, combining three container
types (arrays, bitmaps and runs) in a two-level tree could
surpass competitive implementations of other popular formats (Concise,
WAH, EWAH), being up to hundreds of times faster. For analytical applications, where the bitmaps are not
constantly updated, and where we can afford to sort the data prior to
indexing, applying run compression to the Roaring format
is particularly appealing. The new format has been adopted by existing
systems such as  Apache Spark, Apache Kylin and Druid. 

There are many optimizations and variations on Roaring that future
work should explore. We can implement copy-on-write for containers during unions. We can postpone or omit the computation of the cardinality of some containers (as in our lazy unions). 
We could apply run compression to intermediate results as part of
larger computations when it is likely to improve performance.
We could sometimes improve compression and performance with dynamically resized bitmap containers that only cover the necessary range of values. 
Generally, we could exploit other compression opportunities during serialization.
The Lucene~Roaring implementation~\cite{RoaringDocIdSet} uses a container type 
corresponding to a negated array container: it is one of many other possible container types. Further work could review and 
compare various combinations of container types and their impact on applications.

In the future, we plan to make even better use of existing
and upcoming hardware. For example, operations over Roaring bitmaps could be parallelized at the level of the containers. We could
explicitly seek to exploit single-instruction-multiple-data (SIMD) instructions~\cite{simdcompandinter,lemireboytsov2013decoding}. We could adapt Roaring bitmaps to other processor
architectures such as GPUs and Intel's Xeon Phi.

\ack The Roaring project benefited from so many contributions
that it is not possible to provide an exhaustive list.
Among others, we would like to acknowledge software contributions from  
J.~Alvarado, S.~Chambi, W.~Glynn, R.~Graves, T. Gruben,  B.~Ivanov, T.~Maly, E.~Murphy,  S.~Pellegrino, B.~Potter, G.~Punt\'i, B.~Sperber. We are also grateful to the Druid developers 
(including C.~Allen and F.~Yang) for their help and feedback,
and would like to thank  X.~L\'eaut\'e specifically for providing test cases that motivated our work on run containers. The Apache~Spark
developers (including D.~Liu, S.~Owen, I.~Rashid, R.~Xin and K.~Yao) and Apache~Kylin developers (including L.~Han, L. Yang) provided feedback and a critical assessment. M. Jaffee  helped us correct Algorithm~\ref{algo:set}.

\clearpage
\bibliographystyle{wileyj}
\bibliography{bib/RoaringBitmap}

\appendix

\section{Detailed Container Statistics}
\label{app:detailedstat}
Tables~\ref{tab:contstat} and~\ref{tab:contstatrun} provide detailed container statistics for each dataset for the original Roaring and for Roaring+Run. In each case, the tables indicate how many containers of each type can be found, the total cardinality stored in each container type and the total storage usage (in bytes) of each container type. We provide both the absolute values and the relative values in percentage (e.g., there might be 582 array containers accounting for \SI{75.8}{\percent} of all containers).

In both formats (Roaring without runs and Roaring+Run), 
most containers in \CensInc{} and \Weather{} are array containers even though bitmap containers account for most of the cardinality.
For the sorted datasets in the Roaring+Run format (Table~\ref{tab:contstatrun}), we find no bitmap container whereas run containers account for most of the cardinality and size in bytes.
For the \Censeighten{} dataset, even though it is not sorted, we find no bitmap container after applying run optimization---it is a case where run optimization is maybe not worthwhile as little storage is saved.
\begin{table}
\captionsetup[subfloat]{farskip=2pt,captionskip=1pt}
\centering
\caption{\label{tab:contstat}Container statistics for Roaring without runs.}
\subfloat[\CensInc]{\small
\begin{minipage}{16em}
\begin{tabular}{c|cc}\toprule
& bitmap & array  \\ \midrule
container count & 186             &   582                 \\
&(\SI{24.2}{\percent}) & (\SI{75.8}{\percent})\\
cardinality & \num{6501835}          &   \num{420186}               \\
& (\SI{93.9}{\percent})  & (\SI{6.1}{\percent}) \\
size in bytes & \num{1523712}                 &   \num{841536}       
\\ 
&  (\SI{64.4}{\percent})  & (\SI{35.6}{\percent}) \\
\bottomrule
\end{tabular}
\end{minipage}
}
\subfloat[\CensIncsrt]{\small
\begin{minipage}{10em}
\begin{tabular}{c|cc}\toprule
& bitmap & array  \\ \midrule
 &   186             &   505              \\
 & (\SI{26.9}{\percent}) & (\SI{73.1}{\percent}) \\
 &    \num{5720763}           &   \num{372101}            \\
 & (\SI{93.9}{\percent}) & (\SI{6.1}{\percent})\\
 &   \num{1523712}                  &   \num{745212}    \\
 &   (\SI{67.2}{\percent})                 &  (\SI{32.8}{\percent})   \\
 \bottomrule
\end{tabular}
\end{minipage}
}

\subfloat[\Censeighten]{\small
\begin{minipage}{16em}
\begin{tabular}{c|cc}\toprule
& bitmap & array  \\ \midrule
container count &   5               &   \num{1459}                     \\
 & (\SI{0.3}{\percent})  & (\SI{99.7}{\percent}) \\
cardinality &     \num{28757}           &   \num{975104}         \\
 & (\SI{2.9}{\percent})     &  (\SI{97.1}{\percent})  \\
size in bytes &  \num{40960}            &   \num{1953126}     \\
 & (\SI{2.1}{\percent})  &  (\SI{97.9}{\percent})\\\bottomrule
\end{tabular}
\end{minipage}
}
\subfloat[\Censeightensrt]{\small
\begin{minipage}{10em}
\begin{tabular}{c|cc}\toprule
& bitmap & array  \\ \midrule
 &     16        &   \num{2522}      \\
  & (\SI{0.6}{\percent})        &  (\SI{99.4}{\percent})         \\
 &   \num{498113}          &   \num{182680}            \\
  &  (\SI{73.2}{\percent})   &  (\SI{26.8}{\percent})    \\
 &   \num{131072}      &   \num{370404}  \\
  & (\SI{26.1}{\percent})      &  (\SI{73.9}{\percent})  \\\bottomrule
\end{tabular}
\end{minipage}
}

\subfloat[\Weather]{\small
\begin{minipage}{16em}
\begin{tabular}{c|cc}\toprule
& bitmap & array  \\ \midrule
container count &   575         &   \num{2281}           \\
 &  (\SI{20.1}{\percent})    & (\SI{79.9}{\percent})          \\
cardinality &     \num{10862491}        &   \num{2008136}        \\
 &  (\SI{84.4}{\percent})   & (\SI{15.6}{\percent})     \\
size in bytes &   \num{4710400}                 &   \num{4020834}  \\
 &  (\SI{53.9}{\percent}) & (\SI{46.1}{\percent})   \\\bottomrule
\end{tabular}
\end{minipage}
}
\subfloat[\Weathersrt]{\small
\begin{minipage}{10em}
\begin{tabular}{c|cc}\toprule
& bitmap & array  \\ \midrule
 &    604  &   \num{1564}              \\
  & (\SI{27.9}{\percent})            & (\SI{72.1}{\percent}) \\
&  \num{15325268}         &   \num{782826}            \\
 &  (\SI{95.1}{\percent})  & (\SI{4.9}{\percent})  \\
 &  \num{4947968}           &   \num{1568780}    \\
  & (\SI{75.9}{\percent})        &  (\SI{24.1}{\percent}) \\\bottomrule
\end{tabular}
\end{minipage}
}

\subfloat[\Wikileaks]{\small
\begin{minipage}{16em}
\begin{tabular}{c|cc}\toprule
& bitmap & array  \\ \midrule
container count &   0      &   \num{1892 }           \\
 &  (\SI{0}{\percent})            & (\SI{100}{\percent})      \\
cardinality &     0           &   \num{275355}        \\
 &  (\SI{0}{\percent})  &   (\SI{100}{\percent})  \\
size in bytes &   0          &   \num{554494}   \\
 & (\SI{0}{\percent})  &  (\SI{100}{\percent}) \\\bottomrule
\end{tabular}
\end{minipage}
}
\subfloat[\Wikileakssrt]{\small
\begin{minipage}{10em}
\begin{tabular}{c|cc}\toprule
& bitmap & array  \\ \midrule
 &    18   &   \num{1557}          \\
  &  (\SI{1.1}{\percent})            &  (\SI{98.9}{\percent})    \\
 &   \num{176703}          &   \num{111310}              \\
  &  (\SI{61.4}{\percent})   &  (\SI{38.6}{\percent})  \\
 &  \num{147456}         &   \num{225734}   \\
  & (\SI{39.5}{\percent})   &  (\SI{60.5}{\percent})  \\\bottomrule
\end{tabular}
\end{minipage}
}

\end{table}

\begin{table}
\captionsetup[subfloat]{farskip=2pt,captionskip=1pt}
\centering
\caption{\label{tab:contstatrun}Container statistics for Roaring with runs (Roaring+Run).}
\subfloat[\CensInc]{\small
\begin{minipage}{20em}
\begin{tabular}{c|ccc}\toprule
& bitmap & array & run \\ \midrule
container count & 180          &   553            &   35                 \\
&(\SI{23.4}{\percent})    & (\SI{72}{\percent}) & (\SI{4.6}{\percent}) \\
cardinality & \num{6112509}            &   \num{352778}                &   \num{456734}               \\
&(\SI{88.3}{\percent})  & (\SI{5.1}{\percent}) &  (\SI{6.6}{\percent})\\
size in bytes & \num{1474560}                  &   \num{706662}               &   \num{58938}       
\\ 
&  (\SI{65.8}{\percent})  & (\SI{31.5}{\percent}) & (\SI{2.6}{\percent}) \\
\bottomrule
\end{tabular}
\end{minipage}
}
\subfloat[\CensIncsrt]{\small
\begin{minipage}{10em}
\begin{tabular}{c|ccc}\toprule
& bitmap & array & run \\ \midrule
 &  3              &   277                &   411             \\
 & (\SI{0.4}{\percent})   & (\SI{40.1}{\percent})   &(\SI{59.5}{\percent}) \\
 &   \num{16885}             &   \num{74835}                 &   \num{6001144}         \\
 &  (\SI{0.3}{\percent})  & (\SI{1.2}{\percent}) & (\SI{98.5}{\percent})     \\
 &  \num{24576}           &   \num{150224}              &   \num{275298}  \\
 &    (\SI{5.5}{\percent})                 &  (\SI{33.4}{\percent})  & (\SI{61.2}{\percent})   \\
 \bottomrule
\end{tabular}
\end{minipage}
}

\subfloat[\Censeighten]{\small
\begin{minipage}{20em}
\begin{tabular}{c|ccc}\toprule
& bitmap & array & run \\ \midrule
container count &  0                 &   \num{1332}                   &   132                    \\
 & (\SI{0}{\percent})   & (\SI{91}{\percent}) & (\SI{9}{\percent}) \\
cardinality &  0             &   \num{936719}             &   \num{67142}   \\
& (\SI{0}{\percent}) &  (\SI{93.3}{\percent})  & (\SI{6.7}{\percent}) \\
size in bytes & 0           &   \num{1876102}             &   \num{5696}     \\
 &  (\SI{0}{\percent}) &  (\SI{99.7}{\percent}) &(\SI{0.3}{\percent})\\\bottomrule
\end{tabular}
\end{minipage}
}
\subfloat[\Censeightensrt]{\small
\begin{minipage}{10em}
\begin{tabular}{c|ccc}\toprule
& bitmap & array & run \\ \midrule
 &     0                  &   \num{1061}                 &   \num{1477}     \\
  & (\SI{0}{\percent})        & (\SI{41.8}{\percent}) &    (\SI{58.2}{\percent})       \\
 &  0            &   \num{24871}                 &   \num{655922}           \\
  &  (\SI{0}{\percent})    &  (\SI{3.7}{\percent}) & (\SI{96.3}{\percent})   \\
 &   0          &   \num{51864}               &   \num{112442}  \\
  & (\SI{0}{\percent})      &  (\SI{31.6}{\percent}) &   (\SI{68.4}{\percent}) \\\bottomrule
\end{tabular}
\end{minipage}
}

\subfloat[\Weather]{\small
\begin{minipage}{20em}
\begin{tabular}{c|ccc}\toprule
& bitmap & array & run \\ \midrule
container count &  561             &   \num{2274}               &   21          \\
 &  (\SI{19.6}{\percent})    & (\SI{79.6}{\percent})   &    (\SI{0.7}{\percent})     \\
cardinality &  \num{10371678}          &   \num{1988347}            &   \num{510602}      \\
 &  (\SI{80.6}{\percent})  & (\SI{15.4}{\percent})   &   (\SI{4}{\percent})    \\
size in bytes &  \num{4595712}               &   \num{3981242}              &   \num{60006} \\
 &   (\SI{53.2}{\percent})   & (\SI{46.1}{\percent}) &   (\SI{0.7}{\percent})\\\bottomrule
\end{tabular}
\end{minipage}
}
\subfloat[\Weathersrt]{\small
\begin{minipage}{10em}
\begin{tabular}{c|ccc}\toprule
& bitmap & array & run \\ \midrule
 &    0                 &   909                  &   \num{1259}             \\
  & (\SI{0}{\percent})            & (\SI{41.9}{\percent}) & (\SI{58.1}{\percent})\\
&  0            &   \num{77088}                &   \num{16031006}            \\
 &  (\SI{0}{\percent})   & (\SI{0.5}{\percent})  &  (\SI{99.5}{\percent}) \\
 &  0           &   \num{155994}               &   \num{512186}    \\
  & (\SI{0}{\percent})       &  (\SI{23.3}{\percent}) &  (\SI{76.7}{\percent})\\\bottomrule
\end{tabular}
\end{minipage}
}

\subfloat[\Wikileaks]{\small
\begin{minipage}{20em}
\begin{tabular}{c|ccc}\toprule
& bitmap & array & run \\ \midrule
container count &  0                 &   199                &   \num{1693}          \\
 & (\SI{0}{\percent})            & (\SI{10.5}{\percent})  &    (\SI{89.5}{\percent})   \\
cardinality &   0            &   \num{6377}                 &   \num{268978}     \\
 &  (\SI{0}{\percent})   &   (\SI{2.3}{\percent})  &  (\SI{97.7}{\percent})   \\
size in bytes &  0           &   \num{13152}               &   \num{173770}   \\
 & (\SI{0}{\percent}) &  (\SI{7}{\percent})     & (\SI{93}{\percent})  \\\bottomrule
\end{tabular}
\end{minipage}
}
\subfloat[\Wikileakssrt]{\small
\begin{minipage}{10em}
\begin{tabular}{c|ccc}\toprule
& bitmap & array & run \\ \midrule
 &   0                 &   177                 &   \num{1398}          \\
  &  (\SI{0}{\percent})           &  (\SI{11.2}{\percent})  &   (\SI{88.8}{\percent})  \\
 &   0            &   \num{9352}                &   \num{278661}              \\
  &  (\SI{0}{\percent})   &  (\SI{3.2}{\percent})  &  (\SI{96.8}{\percent}) \\
 & 0           &   \num{19058}                &   \num{26404}   \\
  & (\SI{0}{\percent})   & (\SI{41.9}{\percent}) &  (\SI{58.1}{\percent})\\\bottomrule
\end{tabular}
\end{minipage}
}

\end{table}

\end{document}